\documentclass{elsart}
\usepackage{graphicx}
\begin{document}
\runauthor{B. Fuchs and S. Phleps}
\begin{frontmatter}
\title{Comment on ``General Relativity Resolves Galactic Rotation Without   
   Exotic Dark Matter'' by F.I.~Cooperstock \& S.~Tieu}
\author[Heidelberg]{{Burkhard Fuchs\corauthref{cor}}}
\corauth[cor]{Corresponding author.}
\ead{fuchs@ari.uni-heidelberg.de}
\author[Garching]{Stefanie Phleps}
\address[Heidelberg]{Astronomisches Rechen--Institut am Zentrum f\"ur 
         Astronomie der Universit\"at Heidelberg, M\"onchhofstrasse 12--14, 
	 69120 Heidelberg, Germany }
\address[Garching]{Max--Planck--Institut f\"ur extraterrestrische Physik,
               Giessenbachstrasse, 85748 Garching, Germany}

\begin{abstract}
The general relativistic model of Cooperstock \& Tieu, which attempts to fit
rotation curves of spiral galaxies without invoking dark matter, is tested 
empirically using observations of the Milky Way. In particular, predictions 
for the mass density in the solar neighbourhood and the vertical density 
distribution at the position of the Sun are compared with observations. It 
is shown that the model of Cooperstock \& Tieu, which was so constructed that
it gives an excellent fit of the observed rotation curve, singularly fails to 
reproduce the observed local mass density and the vertical density profile
of the Milky Way.
\end{abstract}
\end{frontmatter}

\section{Introduction}

Recently Cooperstock \& Tieu \cite{ref1} (hereafter CT05) have proposed a new 
approach to the interpretation of rotation curves of spiral galaxies, which is 
based on the theory of general relativity. They argue that even in the case of
such weak
gravitational fields as in galaxies certain non--linear terms in Einstein's
field equations play an important albeit hitherto neglected role. Their 
formalism is applied to concrete examples, and CT05 provide quantitative fits
of the rotation curves of the Milky Way and three further external spiral 
galaxies and they derive mass models for these galaxies. The resulting models 
are quite flattened and their total masses are typically one order of magnitude
lower than those of current models of spiral galaxies. In these models the flat
outer rotation curves are usually modelled by massive dark halos. The low total 
masses estimated by CT05 can be accounted for by the baryonic mass content of 
the galaxies alone. CT05 conclude that it is thus not necessary to invoke 
``exotic dark matter'' to model galactic rotation curves.

Although not yet in print, this spectacular result raised considerable interest
but was also met with scepticism in the astronomical community. For instance
CT05 have not dealt with the dark matter problem of galaxy clusters. A 
conceptual
problem arises from the non continuously differentiable shapes of the density
cusps of the vertical density profiles of the models at the galactic midplanes.
This seems to indicate that each galaxy would at least formally harbour at its
midplane a sheet of negative mass density \cite{ref2}, \cite{ref3}.
Other formal inconsistencies are discussed in \cite{ref4}. In a rebuttal to 
these criticisms CT05 \cite{ref5} maintain the claim of their original paper.

In this {\em comment} we demonstrate how observations of the Milky Way can be
used as an {\em empirical} counter example against CT05's conjecture of the
dynamics of galactic disks. 

\section{The mass density in the solar neighbourhood and the 
vertical mass density profile of the Milky Way at the position of the Sun}

According to CT05's formalism the distribution of mass in their galaxy models is
given by
\begin{eqnarray}
&&
\rho(r,z)=8.36\cdot10^5\,\Bigg( \left( 
\sum_{n=1}^{10}\,k_{\rm n}^2\,C_{\rm n}\,e^{-k_{\rm n}|z|}\,
J_0(k_{\rm n}\,r)\right)^2 \nonumber \\ &&
+\left( \sum_{n=1}^{10}\,k_{\rm n}^2\,C_{\rm n}\,e^{-k_{\rm n}|z|}\,
J_1(k_{\rm n}\,r)\right)^2 \Bigg)\, \frac{M_\odot}{pc^3}\,,
\label{fo1}
\end{eqnarray}
where $J_{0,1}$ denote Bessel functions of the first kind. The coefficients 
$k_{\rm n}$ and $C_{\rm n}$ have been determined by CT05 by fitting the
corresponding model rotation curve to the observed rotation curve of the Milky
Way and are given in their Table 1. Fig.~1 shows in the left panel the vertical 
mass density profile at the position of the Sun, $\rho(r_\odot,z)$, calculated
with Eq.~(1). The Sun lies close to the Galactic midplane, 
$z \approx 0$, and the galactocentric distance of the Sun is about $r_\odot$ =
8 kpc \cite{ref6}, but other determinations are discussed in the literature as
well. Thus density profiles assuming
$r_\odot$ = 7 kpc and $r_\odot$ = 8.5 kpc, which bracket the literature values 
for $r_\odot$, are also shown in Fig.~1. Holmberg \& Flynn \cite{ref7} have 
meticulously compiled an inventory of the contributions by the various phases 
of the interstellar gas and the stellar populations to the mass budget in the 
vicinity of the Sun and find a local mass density of
$\rho(r_\odot,0)$ = 0.094 $M_\odot$/pc$^3$ = 6.3$\cdot 10^{-21}$ kg/m$^3$. 
As described in \cite{ref7} this
value is consistent with dynamical measurements of the local mass
density, if the gravitational force field is calculated in Newtonian
approximation. However, as can be seen from Fig.~1 the mass model of CT05
predicts at the position of the Sun a density of about $\rho(r_\odot,0)$ = 
0.015 $M_\odot$/pc$^3$ = 1.0$\cdot 10^{-21}$ kg/m$^3$. This amounts to only 16
percent of the mass density actually observed in the form of baryons in the 
solar neighbourhood. 

\begin{figure}
\centering
\includegraphics[width=14cm]{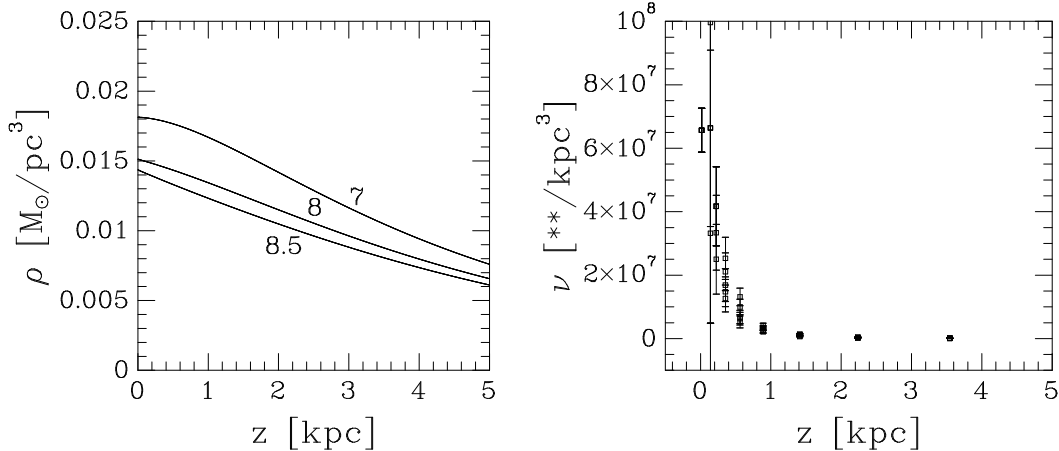}
\caption[]{Predicted versus observed vertical distribution of the mass density 
in the Milky Way at the position of the Sun. Left panel: Vertical distribution
predicted by the mass model of Cooperstock \& Tieu. The profiles are labelled by
the assumed galactocentric distance of the Sun ranging from 7 to 8.5 kpc. Right
panel: The observed distribution of stars perpendicular to the Galactic
midplane.}
\label{fig:1}       % Give a unique label
\end{figure}

Moreover, the predicted shape of the vertical
density distribution looks totally different from what is actually observed. In
the right panel of Fig.~1 the observed number density distribution of stars
perpendicular to the Galactic midplane at the position of the Sun,
$\nu(r_\odot,z)$, is shown. The number densities have been determined with 
counts of K and M stars in five fields of the {\sf Calar Alto Deep Imaging 
Survey} \cite{ref8}. Since the {\sf CADIS} star counts suffer from severe 
Poisson errors near to the midplane due to the conical counting volumes 
(cf.~Fig.~1), the local normalization has been determined by counting stars of
the same spectral types in the {\sf Fourth Catalogue of Nearby Stars} 
\cite{ref8}, \cite{ref9}. The {\sf CADIS} fields point towards different
galactic longitudes and latitudes so that the scatter of the data
points in the right panel of Fig.~1 reflects also some mild variations of the 
vertical shape of the Galactic disk seen in the various direction. We may add
that the vertical density profile derived from {\sf CADIS} data is in perfect
agreement with the results of Zheng et al.~\cite{ref10}. Early type
stars and most of the interstellar gas
are distributed in a narrow layer at the Galactic midplane so that the overall 
distribution of baryons is even more concentrated towards the midplane than 
the late type stars stars, whereas the vertical distribution predicted by 
CT05's model is extremely shallow compared to the observations. Indeed the
implied surface density of the disk at the position of the Sun is 179 
$M_\odot$/pc$^2$ = 0.37 kg/m$^2$. Although the midplane density is much too low,
the predicted surface density is a factor of about four higher than the observed
surface density of baryons of 48 $M_\odot$/pc$^2$ = 0.1 kg/m$^2$ \cite{ref7}. As
can be seen from Eq.~(1) and Eq.~(18) of CT05 any attempt to rescale the model
by increasing the coefficients $k_{\rm n}$ in order to obtain a smaller scale
height would alter also the radial shape of the predicted rotation curve
$V(r,z=0)$ and thus destroy the fit to the observed rotation curve.

This implies that the model of CT05 for the Milky Way, which was so 
constructed
that it gives an excellent fit of the observed rotation curve, has singularly 
failed to reproduce the independent observations of the local Galactic mass 
density and its vertical distribution. This one counter example casts, in our 
view, severe doubts on the viability of Cooperstock \& Tieu's theory of the 
dynamics of galactic disks in general.

\end{document}